\documentclass{article}

\usepackage{PRIMEarxiv}

\usepackage[utf8]{inputenc} % allow utf-8 input
\usepackage[T1]{fontenc}    % use 8-bit T1 fonts
\usepackage{hyperref}       % hyperlinks
\usepackage{url}            % simple URL typesetting
\usepackage{booktabs}       % professional-quality tables
\usepackage{amsfonts}       % blackboard math symbols
\usepackage{nicefrac}       % compact symbols for 1/2, etc.
\usepackage{microtype}      % microtypography
\usepackage{lipsum}
\usepackage{fancyhdr}       % header
\usepackage{graphicx}       % graphics
\graphicspath{{media/}}     % organize your images and other figures under media/ folder
\usepackage{natbib}

\title{Un análisis bibliométrico de la producción científica acerca del agrupamiento de trayectorias GPS
}
\author{
  Gary Reyes \\
  Facultad de Ciencias Matemáticas y Físicas \\
  Universidad de Guayaquil \\
  Guayaquil (Ecuador)\\
  \texttt{gary.reyesz@ug.edu.ec} \\
  \And
  Laura Lanzarini, César Estrebou  \\
  Universidad Nacional de La Plata, Facultad de Informática,   \\
  Instituto de Investigación en Informática LIDI (Centro CICPBA) \\
  Buenos Aires (Argentina)\\
  \texttt{\{laural, cesarest\}@lidi.info.unlp.edu.ar} \\
  \AND
  Aurelio F. Bariviera \\
  Universitat Rovira i Virgili\\
  Department of Business \\
  Reus (Spain) \\
  \texttt{aurelio.fernandez@urv.cat} \\
}

\begin{document}
\maketitle

\begin{abstract}
Los algoritmos o métodos de agrupamiento para trayectorias GPS se encuentran en constante evolución debido al interés que despierta en parte de la comunidad científica. 
Con el desarrollo de los algoritmos de agrupamiento considerados tradicionales han surgido mejoras a estos algoritmos e incluso métodos únicos considerados como ``novedad'' para la ciencia. 
Este trabajo tiene como objetivo analizar la producción científica que existe alrededor del tema ``agrupamiento de trayectorias GPS'' mediante la bibliometría. 
Por lo tanto, fueron analizados un total de 559 artículos de la colección principal de Scopus, realizando previamente un filtrado de la muestra generada para descartar todo aquel artículo que no tenga una relación directa con el tema a analizar. 
Este análisis establece un ambiente ideal para otras disciplinas e investigadores, ya que entrega un estado actual de la tendencia que lleva la temática de estudio en su campo de investigación. 
\end{abstract}

% keywords can be removed
% \keywords{Trajectory clustering \and GPS trajectories \and Trajectory clustering algorithms \and Bibliometry}

\section{Introducción}

El estudio bibliométrico es una disciplina que ha tenido un crecimiento importante dentro de la comunidad científica en los últimos años. Eugene Garfield, con el establecimiento del Instituto de Información Científica (ISI) en la década de 1960, inició la medición de artículos, revistas, investigadores e instituciones \citep{merediz-sola_bibliometric_2019}. La investigación bibliométrica examina la autoría, la publicación, las citas y el contenido aplicando medidas cuantitativas a un cuerpo (corpus) de literatura \citep{haddow_bibliometric_2018}. 
En la actualidad los artículos científicos se almacenan e indexan en grandes bases de datos científicas, permitiendo medir los parámetros que tienen, como sus palabras claves, números de citas, números de autores, colaboración e impacto del autor, la producción científica anual, entre otros. La idea principal es que al conseguir más citas en un campo científico indica mayor importancia, calidad y es más destacable \citep{dede_mapping_2022}.
La razón para indexar artículos está dada por lo siguiente: los autores citan otros trabajos por su idea central, esto es debido a la conexión que tienen con el tema central de su investigación o trabajo. Dado que cualquier autor puede seleccionar que artículo va a citar, incluyendo solamente los más relevantes y relacionados con su artículo, la mayoría de los artículos que se citan podrían demostrar el impacto o la importancia que han tenido dentro de su campo científico. La información que se puede obtener puede ser aprovechada por diversas instituciones, ya que se otorga información valiosa tanto sobre el impacto individual como el agregado. Por lo tanto, podría ayudar en la contratación de maestros o en idear estrategias de investigación en las universidades y los consejos de investigación, sin embargo, los estudios bibliométricos también pueden ayudar con información acerca de la historia que ha tenido cierta temática, además, de dar a conocer el alcance o la tendencia que lleva dicho tema de investigación. De esta forma se ayuda a los nuevos investigadores a tener una idea sobre el impacto que tiene un tema de investigación en su campo científico \citep{singh_bibliometric_2023}. 
Este tipo de análisis se hace posible mediante la disponibilidad de grandes bases de datos bibliográficas como Scopus o Web of Science, entre otros. Estos servicios de indexación son un medio importante para el proceso de evaluación en el ámbito académico. 

Scopus es una base de datos bibliográfica que recopila citas y resúmenes de una amplia variedad de fuentes neutrales. Estos recursos son cuidadosamente seleccionado por expertos independientes, quienes son líderes reconocidos en sus respectivos campos disciplinarios. 
Scopus ofrece a los investigadores una gama de herramientas de descubrimiento y análisis. Esta plataforma no solo facilita la búsqueda y recuperación de información relevante, sino que también promueve la colaboración y el intercambio de ideas entre individuos e instituciones en la comunidad científica. Con un amplio alcance, Scopus indexa contenido proveniente de más de 7000 editores, abarcando una diversidad de disciplinas. Además, alberga una vasta colección de datos, con más de 91 millones de registros, incluyendo más de 94 000 perfiles de afiliación y la contribución de más de 17 millones de autores. 

Desde un nivel macroscópico, se pueden determinar métricas que son comunes con muchas revistas y son útiles para diferentes interesados. Sin embargo, algunas características cambian de un contexto o disciplina a otra. Existe una cantidad de investigadores y de revistas que actúan de forma desigual. En los últimos años se ha producido una expansión en el número de revistas y un aumento en los periodos en que se publican, esto puede ser gracias a la expansión del sector académico en varios países, aumentando paulatinamente en la última década en diversos países. Además, las disciplinas científicas tienen diferentes parámetros con respecto a la publicación de un artículo. Por lo tanto, es importante estudiar, sus características y/o temas equivalentes, con el fin de proporcionar una clasificación significativa para los parámetros bibliométricos.  

El objetivo de este artículo es analizar la metadata de todos los artículos indexados en la base de datos bibliográfica Scopus que realizan ``algoritmos o métodos para agrupamiento de trayectorias GPS''. Se destaca también, que las muestras generadas por la base de datos bibliográfica fueron filtradas manualmente para excluir todos los artículos que no sean parte del campo de estudio. El presente artículo proporcionará información útil sobre las principales revistas que están interesadas en publicar artículos acerca de este tema en particular, así como la evolución que ha tenido su campo científico con el paso del tiempo. Además, se discuten otros aspectos como los autores más citados, las áreas en que más se publican estos artículos, el número de publicaciones por año, los diagramas estratégicos sobre el impacto de los temas, la evolución temática, entre otros. 

El análisis bibliométrico viene dado gráficamente por el software VOSviewer que es una herramienta de software para crear mapas basados en datos de red, para visualizar y explorar estos mapas \citep{van_eck_software_2010}, entre ellos los gráficos de citas, fuentes y autores. Además, se hace uso del paquete bibliometrix y su interfaz gráfica biblioshiny del lenguaje de programación R, que fue desarrollado por \citet{aria_bibliometrix_2017} para realizar el análisis acerca de la distribución gráfica del autor correspondiente, los artículos más citados, las palabras claves principales, las principales fuentes de publicación, los diagramas estratégicos de las palabras claves y la evolución temática de las palabras claves. Ambos softwares son de código libre, lo que permite al investigador utilizar todas sus funcionalidades, como el artículo más citado, la coautoría, entre otros. 
El resto del documento está estructurado de la siguiente manera. La sección 2 describe la literatura sobre los estudios bibliométricos o semejantes que se hayan realizados con respecto al tema. La sección 3 detalla los datos bajo el análisis, además, de los principales hallazgos del estudio por medio de bibliometrix y su interfaz gráfica biblioshiny. En la sección 4 se realiza el análisis por medio de VosViewer de los indicadores seleccionados. En la sección 5 se explica las posibles líneas de investigación que pueden desprenderse del análisis. Finalmente, la sección 6 extrae las principales conclusiones.

\section{Revisión de literatura}

La bibliometría es un campo de investigación dentro de las ciencias bibliotecarias y de la información que estudia el material bibliográfico utilizando métodos cuantitativos \citep{donthu_how_2021}. Con el paso de los años, la bibliometría se ha convertido en un campo de interés para clasificar y desarrollar resúmenes que representen los principales resultados de una temática en específico. Existen estudios bibliométricos con diversos temas en diferentes campos científicos. Por ejemplo, \citet{liu_research_2019} realizan un análisis bibliométrico acerca de la investigación de big data. \citet{derudder_shifting_2019} realizan un análisis bibliométrico sobre la posición de JTRG (Journal of Transport Geography) en las diferentes agendas o campos de investigación con respecto a las ciencias geográficas y del transporte. Por otra parte, \citet{zhang_global_2020} realizan un análisis estadístico en profundidad del conocimiento global de la investigación de microplásticos. \citet{ramona_bitcoin_2019} desarrollan un estudio bibliométrico para evaluar la literatura sobre bitcoins a partir de las estructuras y redes de las ciencias publicadas entre 2012 y 2019. \citet{shen_bibliometric_2022} realizan un análisis bibliométrico acerca del enrutamiento de vehículos de distribución de material de emergencia. \citet{azam_autonomous_2022} realizan un análisis bibliométrico para mapear el estado del arte de la producción científica sobre vehículos autónomos en condiciones de tráfico mixto.
Con respecto al tema de estudio, no se identificaron artículos que realicen estudios o análisis bibliométricos. Sin embargo, se menciona un artículo como el de \citet{yuan_review_2017} que realiza un análisis general de los algoritmos de agrupamiento más representativos que fueron desarrollados, destacando las ventajas y desventajas de cada uno, sus medidas de similitud, resume criterios sobre los resultados de los agrupamientos y por último comenta acerca de los escenarios en que pueden aplicarse. 
El presente artículo puede convertirse en un punto de partida para los nuevos investigadores que ingresen a este campo, otorgándoles una visón general y completa de las tendencias actuales que existen en torno a los algoritmos o métodos para agrupar trayectorias GPS, identificando, los principales investigadores, las instituciones y revistas que publican artículos relacionados con este campo de investigación.

\section{Datos y Resultados}

Se trabajó con la metadata bibliográfica (bibliographic metadata) de los artículos que se encuentran indexados en la base de datos bibliográfica Scopus. Por lo tanto, se seleccionaron solamente los artículos que desarrollen o investiguen acerca de la ``agrupamiento de trayectorias GPS''. Scopus alberga un total de 559 documentos de la muestra, publicados en 333 fuentes (revistas, libros, etc), durante el período de 2002-2023. Estos documentos fueron (co) escritos por 1416 personas, la gran mayoría de los documentos son de varios autores, sin embargo, solo 11 documentos son de autoría única. La media de autores por documentos es 3,87.
Al realizar el análisis se observó que se concentraban en dos áreas de investigación principales: Ciencias de la Computación e Ingeniería. Scopus asigna artículos indexados a una o más áreas de investigación. Los 559 artículos de la muestra fueron asignados a diversas áreas de investigación dando un total de 1094, es decir que pertenecen a más de un área en específico. 
Las cinco principales áreas de investigación se muestran en la Tabla~\ref{tab:table_1}.

% Tabla I
\begin{table}
\caption{Principales áreas de investigación asignadas a los trabajos de la muestra. Fuente: Scopus.}
\centering
\begin{tabular}{lll}
\toprule
Áreas de investigación     & Registros     & \% de 1094 \\
\midrule
Ciencias de la computación & 391 & 35,74\% \\
Ingeniería & 176 & 16,09\%\\
Ciencias sociales & 125 & 11,43\% \\
Matemáticas & 123 & 11,24\% \\
Tierra y ciencias planetarias & 69 & 6,31\% \\
Total de las 5 principales áreas de investigación & 884 & 80,80\% \\
\bottomrule
\end{tabular}
\label{tab:table_1}
\end{table}

El detalle de las publicaciones anuales de artículos se muestra en la Tabla~\ref{tab:table_2}. 
En Scopus se observa que los primeros años tuvieron la acogida de pocos artículos relacionados con el ``agrupamiento de trayectorias GPS'', aunque en la última década la cantidad de artículos que se publicaron se han ido incrementado posiblemente por la acogida de la comunidad científica. El total de registros de la muestra cuenta con una tasa de crecimiento promedio por año del 15,6\% desde el 2002 hasta el 2023.

% Tabla II
\begin{table}
\caption{Número de artículos publicados por año. Fuente: Scopus.}
\centering
\begin{tabular}{lll}
\toprule
Años     & Artículos     & Tasa de crecimiento anual \\
\midrule
2002 & 2 & - \\
2003 & 2 & 0,0\% \\
2004 & 1 & -50,00\% \\
2005 & 0 & -100,00\% \\
2006 & 0 & - \\
2007 & 1 & - \\
2008 & 2 & 100,00\% \\
2009 & 11 & 450,00\% \\
2010 & 12 & 9,09\% \\
2011 & 12 & 0,00\% \\
2012 & 15 & 25,00\% \\
2013 & 21 & 40,00\% \\
2014 & 30 & 42,86\% \\
2015 & 31 & 3,33\% \\
2016 & 37 & 19,35\% \\
2017 & 48 & 29,73\% \\
2018 & 55 & 14,58\% \\
2019 & 68 & 23,64\% \\
2020 & 57 & -16,18\% \\
2021 & 52 & -8,77\% \\
2022 & 60 & 15,38\% \\
2023 & 42 & -30,00\% \\
Total & 559 & 15,6\% \\
\bottomrule
\end{tabular}
\label{tab:table_2}
\end{table}

\subsection{Distribución geográfica del autor correspondiente}

La Tabla~\ref{tab:table_3} muestra a China como el principal país cuyos autores han publicado más documentos, seguido por USA como el segundo país que más documentos publicados tiene. Los diez primeros países acumulan el 53,9\% de los artículos publicados relacionados con ``agrupamiento de trayectorias GPS''. Las siglas PSP, PVP y PVP Ratio corresponden a ``Publicaciones de un Solo País'', ``Publicaciones de Varios Países'' y ``Proporción de las Publicaciones de Varios Países''. 
La Tabla~\ref{tab:table_4} muestra los principales países, ordenados por el número total de citas. El promedio de citas de todos los artículos es de 21,92. China y USA, son los dos países con más artículos publicados y citaciones totales, se sitúan por encima de esta cifra, con un promedio de 19,30 y 34,60 respectivamente. A pesar de que China es el primer país en términos de artículos publicados, tiene el segundo promedio más bajo de citas por artículo entre los países líderes. También es importante destacar que USA es el país con el promedio más alto de citas por artículo, lo que puede utilizarse como un denominador común en la importancia científica media o la calidad de los artículos. 
Los países que menos colaboran internacionalmente con otros países son Países Bajos y Tailandia al tener una tasa de publicaciones del 0\%. El país que más colabora internacionalmente con otros países es China, donde el 30,80\% de los trabajos son de este tipo. 

% Tabla III
\begin{table}
\caption{Diez países de autores correspondientes. Fuente: Scopus.}
\centering
\begin{tabular}{llllll}
\toprule
País & Artículos & Frecuencia & PSP & PVP & PVP Ratio\\
\midrule
China       & 203 & 36,3\% & 159 & 44 & 21,7\% \\
USA         & 29 & 5,2\% & 17 & 12 & 41,4\% \\
India       & 16 & 2,9\% & 14 & 2 & 12,5\% \\
Italia      & 13 & 2,3\% & 12 & 1 & 7,7\% \\
Corea       & 11 & 2,0\% & 8 & 3 & 27,3\% \\
Portugal    & 8 & 1,4\% & 4 & 4 & 50,0\% \\
Japón       & 6 & 1,1\% & 5 & 1 & 16,7\% \\
Australia   & 5 & 0,9\% & 3 & 2 & 40,0\% \\
Francia     & 5 & 0,9\% & 2 & 3 & 60,0\% \\
Alemania    & 5 & 0,9\% & 3 & 2 & 40,0\% \\
Total 10 países & 301 & 53,9\% & 227 & 74 & 31,7\% \\
\bottomrule
\end{tabular}
\label{tab:table_3}
\end{table}

% Tabla IV
\begin{table}
\caption{Diez principales citas totales por país. Fuente: Scopus.}
\centering
\begin{tabular}{lll}
\toprule
País & Total de citas & Citas promedio de artículos \\
\midrule
China       & 3908 & 19,30 \\
USA         & 1004 & 34,60 \\
Turquía     & 400 & 400,00 \\
Italia      & 202 & 15,50 \\
Hong Kong   & 173 & 34,60 \\
Suiza       & 173 & 34,60 \\
Grecia      & 167 & 33,40 \\
España      & 156 & 39,00 \\
Francia     & 127 & 25,40 \\
Australia   & 118 & 23,60 \\
Total (Todos los países) & 7138 & 21,92 \\
\bottomrule
\end{tabular}
\label{tab:table_4}
\end{table}

\subsection{Principales fuentes de publicación}

La Tabla~\ref{tab:table_5} muestra las diez fuentes principales que publican artículos relacionados con ``algoritmos de agrupamiento de trayectorias''. Las tres primeras son Lecture Notes in Computer Science (LNCS) (incluida su subserie Lecture Notes en Inteligencia Artificial, LNAI y Apuntes de Conferencias en Bioinformática, LNBI) es una serie de actas de congresos que publica los últimos avances en investigación en todas las áreas de la informática. ISPRS International Journal of Geo-Information que es una revista internacional de acceso abierto revisada por pares sobre geoinformación. IEEE Access que es una importante revista multidisciplinaria de acceso abierto. GIS: Proceedings of the ACM International Symposium On Advances In Geographic Information Systems son Conferencias Internacionales ACM SIGSPATIAL sobre avances en investigaciones interdisciplinarios en todos los aspectos de los sistemas de información geográfica. ACM International Conference Proceeding Series son una serie de Actas de Conferencias Internacionales (ICPS) que proporcionan un mecanismo para publicar los contenidos de conferencias, simposios técnicos y talleres de alta calidad. La International Journal of Geographical Information Science es una revista que permite la revisión por pares, publica temas relacionados con la ciencia de la información geográfica fundamental y computacional entre otros. Cluster Computing-The Journal of Networks Software Tools and Applications es una revista científica revisada por pares sobre procesamiento paralelo, sistemas informáticos distribuidos y redes de comunicación informática. IEEE Transactions on Intelligent Transportation Systems es una revista que se publica por medio de IEEE Access, entre el alcance de los temas que publican son: comunicaciones (intervehículo y de vehículo a camino), computadoras(hardware, software), sistemas de información (bases de datos, fusión de datos, seguridad), entre otros. International Archives Of The Photogrammetry, Remote Sensing And Spatial Information Sciences - Isprs Archives es una serie de archivos revisados por pares actas publicadas por la Sociedad Internacional de Fotogrametría y Teledetección (ISPRS). La revista científica Jiaotong Yunshu Xitong Gongcheng Yu Xinxi/ Journal of Transportation Systems Engineering and Information Technology está incluida en la base de datos Scopus, sus principales áreas temáticas de los artículos publicados son Aplicaciones de Ciencias de la Computación, Ingeniería de Sistemas y Control, Modelado y Simulación, y Transporte. Y finalmente Transactions in GIS es una revista internacional revisada por pares que publica artículos de investigación originales, artículos de revisión y notas técnicas breves sobre los últimos avances y las mejores prácticas en las ciencias espaciales.

% Tabla V
\begin{table}
\caption{Las diez fuentes más relevantes. Fuente: Scopus.}
\centering
\begin{tabular}{lll}
\toprule
Fuentes     & \# Artículos     & Tipo \\
\midrule
\begin{tabular}[c]{@{}l@{}} Lecture Notes in Computer Science (Including Subseries \\Lecture Notes in Artificial Intelligence and Lecture \\Notes in Bioinformatics) \end{tabular} & 38 & Actas de Conferencias \\
\begin{tabular}[c]{@{}l@{}} ISPRS International Journal of Geo-Information \end{tabular} & 17 & Revista \\\begin{tabular}[c]{@{}l@{}} IEEE Access \end{tabular} & 15 & Revista \\
\begin{tabular}[c]{@{}l@{}} GIS: Proceedings of the ACM International Symposium on \\Advances in Geographic Information Systems \end{tabular} & 13 & Actas de Conferencias \\
\begin{tabular}[c]{@{}l@{}} ACM International Conference Proceeding Series \end{tabular} & 12 & Actas de Conferencias \\
\begin{tabular}[c]{@{}l@{}} International Journal of Geographical Information Science \end{tabular} & 11 & Revista \\
\begin{tabular}[c]{@{}l@{}} IEEE Transactions on Intelligent Transportation Systems \end{tabular} & 8 & Revista \\
\begin{tabular}[c]{@{}l@{}} International Archives of the Photogrammetry, Remote \\Sensing and Spatial Information Sciences - ISPRS Archives \end{tabular} & 8 & Revista \\
\begin{tabular}[c]{@{}l@{}} Jiaotong Yunshu Xitong Gongcheng Yu Xinxi/Journal of \\Transportation Systems Engineering and Information Technology \end{tabular} & 7 & Revista \\
\begin{tabular}[c]{@{}l@{}} Transactions in GIS \end{tabular} & 7 & Revista \\
\bottomrule
\end{tabular}
\label{tab:table_5}
\end{table}

\subsection{Artículos más citados}

La Tabla~\ref{tab:table_6} muestra la lista de los 10 principales artículos categorizados como un artículo altamente citado en Scopus. Según \citet{gonzalez-betancor_porcentaje_2015} los trabajos más citados son aquellos que han recibido un número de citas igual o superior que las del percentil $q$ para su campo y año de publicación. Un artículo altamente citado es reconocido por poseer excelencia científica, estableciendo las bases para el campo en que se centra su contexto en el mundo. Por lo tanto, sirven para destacar los artículos importantes de los diferentes campos. Estos artículos se convierten en caminos para la investigación. 
El primer artículo más citado es propuesto por \citet{yuan_t-drive_2010} que diseñaron un enfoque de agrupación basada en varianza-entropía para estimar la distribución del tiempo de viaje entre dos puntos de referencia en diferentes franjas horarias.
\citet{abul_never_2008} proponen un concepto novedoso de k-anonimato basado en la co-localización que explota la incertidumbre inherente del paradero del objeto en movimiento.
\citet{jing_yuan_t-drive_2013} diseñaron un enfoque de agrupamiento basado en varianza-entropía para estimar la distribución del tiempo de viaje entre dos puntos de referencia en diferentes intervalos de tiempo. 
\citet{tang_uncovering_2015} utilizan una matriz observada del área central en la ciudad de Harbin para modelar los patrones de distribución del tráfico basados en método de maximización de la entropía, y el desempeño de la estimación verifica su efectividad.
\citet{schroedl_mining_2004} presentan un enfoque para inducir mapas de alta precisión a partir de trazas de vehículos equipados con receptores GPS diferenciales. 
\citet{guo_discovering_2012} presentan nueva metodología para detectar la ubicación de patrones espaciales y estructuras incrustadas en origen-destino de movimientos.
\citet{abul_anonymization_2010} abordan el problema de la anonimización de bases de datos de objetos en movimiento y proponen un concepto novedoso de k-anonimato basado en la co-localización, que explota la incertidumbre inherente del paradero de objetos en movimiento.
\citet{hutchison_incremental_2010} proponen un marco de agrupamiento incremental para trayectorias que contiene dos partes,  el mantenimiento de microclústeres en línea y la creación de macroclústeres fuera de línea.
\citet{chen_tripimputor_2018} proponen un marco probabilístico para inferir propósitos de viaje, tiene una fase que identifica áreas de actividad y calcula probabilidades usando el teorema de Bayes, mientras que la segunda fase agrupa puntos de entrega y coincide con áreas de actividad para respuestas en tiempo real.
Finalmente, \citet{monreale_movement_2010} presentan un método que garantiza el anonimato en datos de trayectorias mediante una transformación basada en generalización espacial y k-anonimato, proporcionando una protección formal de datos con un límite superior teórico de reidentificación.

% Tabla VI
\begin{table}
\caption{Los diez artículos más citados, ordenados descendentemente por el número de citas. Fuente: Scopus.}
\centering
\begin{tabular}{lll}
\toprule
Autor (año) y Título     & Fuente     & \# Citas \\
\midrule
\begin{tabular}[c]{@{}l@{}} \citet{yuan_t-drive_2010}. T-drive: driving \\directions based  on taxi trajectories.  \end{tabular}  & \begin{tabular}[c]{@{}l@{}} GIS: International Conference on Advances \\in Geographic Information Systems \end{tabular}& 884 \\
\begin{tabular}[c]{@{}l@{}} \citet{abul_never_2008}. Never\\Walk Alone: Uncertainty for Anonymity \\in Moving Objects Databases.  \end{tabular}  & \begin{tabular}[c]{@{}l@{}}2008 IEEE 24th International \\Conference on Data Engineering \end{tabular}& 400 \\
\begin{tabular}[c]{@{}l@{}} \citet{jing_yuan_t-drive_2013}. T-Drive: \\Enhancing Driving Directions with \\Taxi Drivers' Intelligence.  \end{tabular}  & IEEE Xplore & 348 \\
\begin{tabular}[c]{@{}l@{}} \citet{tang_uncovering_2015}. Uncovering urban \\human mobility from large scale taxi\\GPS data.  \end{tabular}  & \begin{tabular}[c]{@{}l@{}}Physica A: Statistical Mechanics and \\its Applications \end{tabular} & 232 \\
\begin{tabular}[c]{@{}l@{}} \citet{schroedl_mining_2004}. Mining GPS \\Traces for Map Refinement.  \end{tabular}  & Data Mining and Knowledge Discovery & 197 \\
\begin{tabular}[c]{@{}l@{}} \citet{guo_discovering_2012}. Discovering \\Spatial Patterns in Origin‐Destination\\Mobility Data.  \end{tabular}  & Transactions in GIS & 145 \\
\begin{tabular}[c]{@{}l@{}} \citet{abul_anonymization_2010}. \\Anonymization of moving objects \\databases by clustering and perturbation.  \end{tabular}  & Information Systems & 144 \\
\begin{tabular}[c]{@{}l@{}} \citet{hutchison_incremental_2010}. Incremental \\Clustering for Trajectories.  \end{tabular}  & Springer Berlin Heidelberg & 132 \\
\begin{tabular}[c]{@{}l@{}} \citet{chen_tripimputor_2018}. TripImputor: \\Real-Time Imputing Taxi Trip Purpose \\Leveraging Multi-Sourced Urban Data.  \end{tabular}  & \begin{tabular}[c]{@{}l@{}}IEEE Transactions on Intelligent \\Transportation Systems \end{tabular} & 125 \\
\begin{tabular}[c]{@{}l@{}} \citet{monreale_movement_2010}. Movement data\\anonymity through generalization.  \end{tabular}  & Transactions on Data Privacy & 125 \\
\bottomrule
\end{tabular}
\label{tab:table_6}
\end{table}

La Tabla~\ref{tab:table_7} muestra los autores más productivos. La tabla fué realizada a partir de una búsqueda manual, ya que, bibliometrix al realizar el análisis del parámetro de los autores, no logra diferenciar entre uno u otro autor que tenga el mismo apellido con la misma inicial de su nombre, por lo tanto, se obtuvieron los siguientes resultados. En primer lugar, se encuentran Wang Haoyu y Li Jinhong con 7 artículos publicados respectivamente, seguidos por Xu Hao y Liu Xintao con 6 artículos, finalmente Li Yanhua con 4 artículos publicados.

% Tabla VII
\begin{table}
\caption{Autores más productivos. Fuente: Scopus.}
\centering
\begin{tabular}{lll}
\toprule
Autores & Institución & \# Artículos \\
\midrule
Wang Haoyu & Yunnan University, Kunming, China & 16 \\
Li Jinhong & North China University of Technology, Beijing, China & 15 \\
Li Xue & Shandong University of Science and Technology, Qingdao, China & 12 \\
Liu Yizhi & Hunan University of Science and Technology, Xiangtan, China & 12 \\
Li Qing & Shandong University of Science and Technology, Qingdao, China & 11 \\
\bottomrule
\end{tabular}
\label{tab:table_7}
\end{table}

\subsection{Palabras claves principales.}

La Tabla~\ref{tab:table_8} muestra las diez palabras clave más utilizadas en los artículos de agrupamiento de trayectorias GPS. Scopus proporciona dos tipos de palabras clave: (a) Author Keywords, que son las proporcionadas por los autores originales, y (b) Keywords-Plus, que son aquellas extraídas de los títulos de las referencias citadas. Las Keyword Plus son generadas automáticamente por un algoritmo informático. Las dos palabras clave de autor más frecuentes son ``clustering'' y ``trajectory''. 
Por su parte las Keywords-Plus en sus primeros lugares contienen las palabras ``trajectories'' y ``clustering algorithms'', presentes en artículos como \citet{reyes_batch_2023}. 
Se observa que al menos cuatro de las principales Keywords en ambos tipos coinciden, posiblemente porque engloban todo lo que tiene que ver con trayectorias y datos GPS, además, de que son utilizados en el proceso de la minería de datos.

% Tabla VIII
\begin{table}
\caption{Palabras clave principales. Fuente: Scopus.}
\centering
\begin{tabular}{llll}
\toprule
Palabras clave del autor & \# Artículos & Palabras clave plus  & \# Artículos \\ %  (DE) % (ID)
\midrule
clustering              & 67 & trajectories	                & 278 \\
trajectory              & 34 & clustering algorithms        & 169 \\
trajectory clustering   & 27 & global positioning system    & 119 \\
gps                     & 26 & data mining                  & 105 \\
gps trajectory          & 24 & taxicabs                     & 88 \\
dbscan                  & 22 & cluster analysis             & 75 \\
data mining             & 21 & roads and streets            & 70 \\
gps data                & 21 & gps trajectories             & 61 \\
gps trajectories        & 20 & trajectory clustering        & 61 \\
big data                & 14 & gps                          & 57 \\
\bottomrule
\end{tabular}
\label{tab:table_8}
\end{table}

\subsection{Diagrama estratégico de las palabras claves.}

En el diagrama estratégico se pueden observan los temas que están emergiendo, son tendencia, están o han desaparecido de un campo de investigación mediante el análisis de las palabras claves.
Cuando se utiliza el análisis de palabras conjuntas para cartografiar la ciencia, se obtienen grupos de palabras clave (y sus interconexiones). Estos grupos se consideran temas. Cada tema de investigación obtenido en este proceso se caracteriza por dos parámetros ``densidad'' y ``centralidad'' \citep{cobo_approach_2011}.
El paquete de bibliometrix mediante su interfaz biblioshiny permite crear el mapa temático o diagrama estratégico de las palabras claves, títulos y resúmenes. Dada la interpretación del diagrama estratégico de \citet{cobo_approach_2011}, el diagrama otorgado por bibliometrix se analiza de la siguiente manera:

\begin{enumerate}
    \item Los temas del cuadrante superior derecho están bien desarrollados y son importantes para la estructuración de un campo de investigación. Se les conoce como los temas motores de la especialidad, dado que presentan una fuerte centralidad y alta densidad. La ubicación de los temas en este cuadrante implica que están relacionados externamente con conceptos aplicables a otros temas que están estrechamente relacionados conceptualmente.
    \item 	Los temas en el cuadrante superior izquierdo tienen vínculos internos bien desarrollados, pero vínculos externos sin importancia y, por lo tanto, son de importancia marginal para el campo. Estos temas son de carácter muy especializado y periférico.
    \item 	Los temas del cuadrante inferior izquierdo están poco desarrollados y son marginales. Los temas de este cuadrante tienen baja densidad y baja centralidad, representando principalmente temas emergentes o desaparecidos.
    \item 	Los temas del cuadrante inferior derecho son importantes para un campo de investigación, pero no están desarrollados. Entonces, este cuadrante agrupa temas básicos transversales y generales.
\end{enumerate}

\begin{figure}
\centering
\includegraphics[width=14cm]{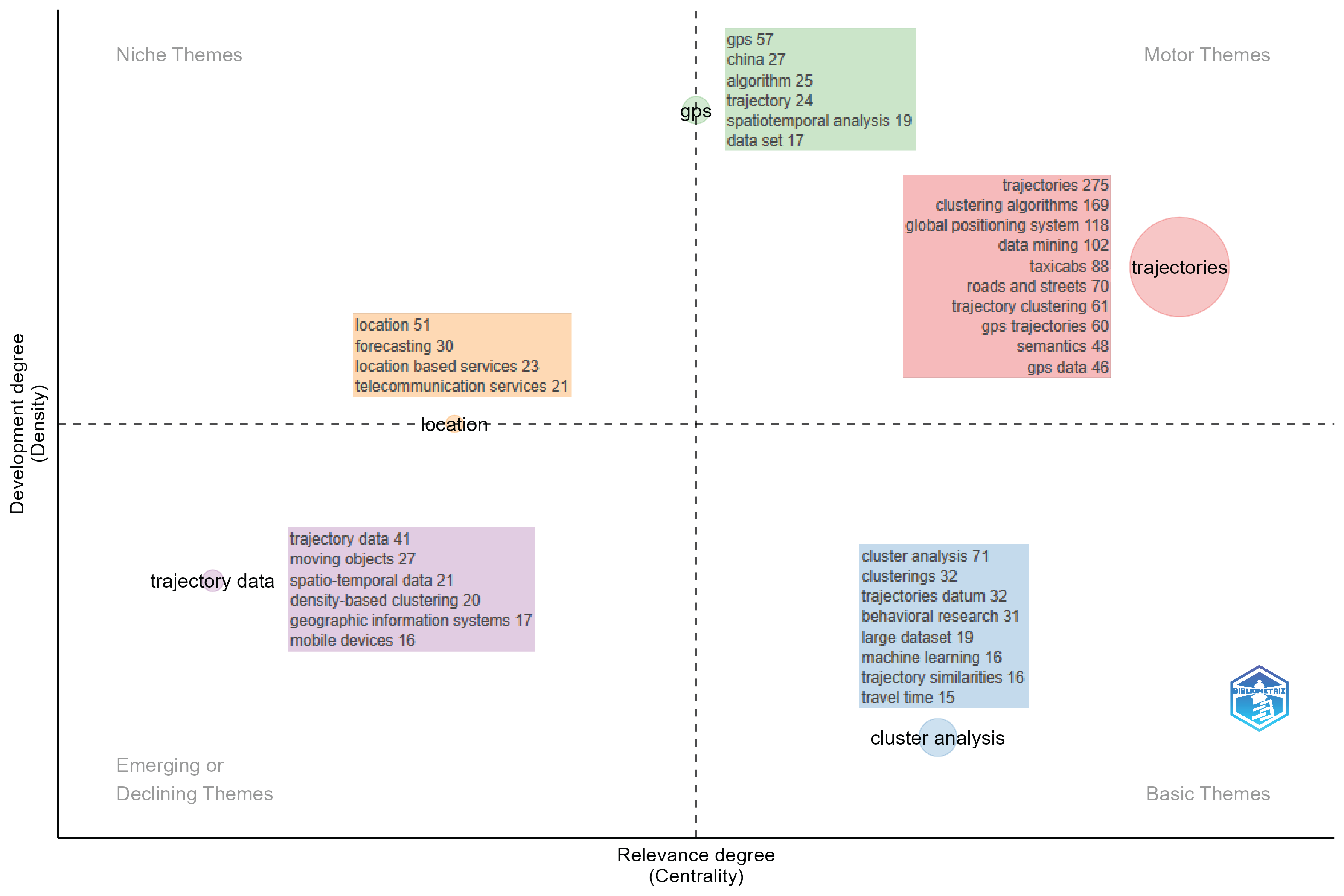}
\caption{Diagrama estratégico de las KeyWords Plus, generado con bibliometrix. Fuente: Scopus.}
\label{fig:fig_1}
\end{figure}

\begin{figure}
\centering
\includegraphics[width=14cm]{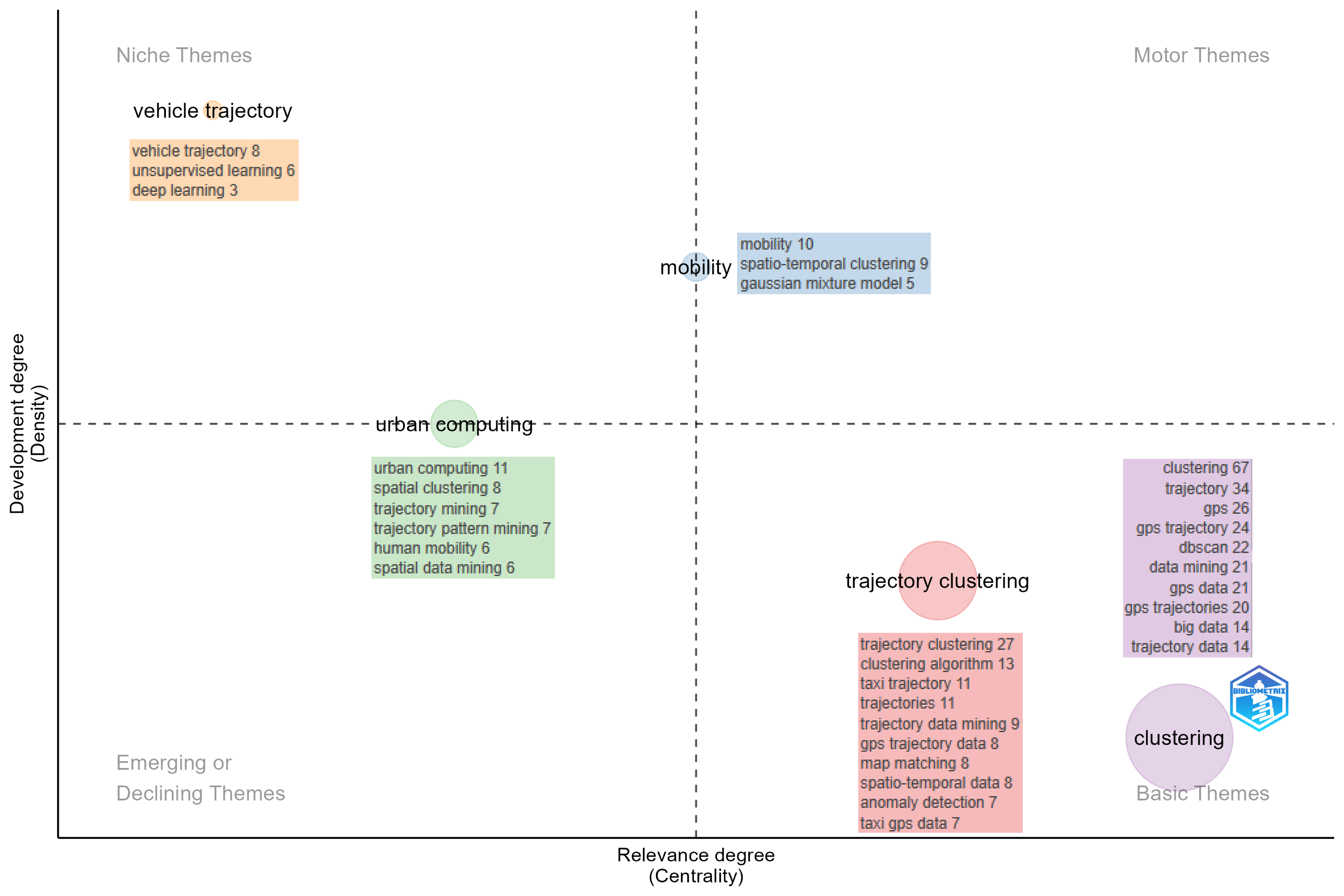}
\caption{Diagrama estratégico de las palabras clave del autor, generado con bibliometrix. Fuente: Scopus.}
\label{fig:fig_2}
\end{figure}

En la Figura~\ref{fig:fig_1} y Figura~\ref{fig:fig_2} se observan los diagramas estratégicos pertenecientes a las KeyWords Plus y las palabras claves del autor de Scopus.
Para la Figura~\ref{fig:fig_1} se muestran las KeyWords Plus, su cuadrante superior derecho contiene los temas ``trajectories'' y parte del tema ``gps'' considerados como un grupo de subtemas bien desarrollados e importantes para el campo de investigación de los ``algoritmos o métodos para agrupamiento de trayectorias GPS''. 
Su cuadrante superior izquierdo cuenta parcialmente con los temas ``gps'' y ``location'', es decir, que contienen subtemas bien desarrollados, aunque no son de importancia para el campo de investigación. Su cuadrante inferior izquierdo contiene la palabra clave ``trajectory data'' y la otra mitad del tema ``gps'' dentro de estos temas se encuentran subtemas que están pocos desarrollados, no son tomados en cuenta, son emergentes o han desaparecidos. 
Por último, su cuadrante inferior derecho cuenta con el tema ``cluster analysis'', es decir, contienen subtemas importantes, aunque no están del todo desarrollados.

%FIGURA 1. Diagrama estratégico de las keyWords Plus, generado con bibliometrix. Fuente: Web of Science.

Para la Figura~\ref{fig:fig_2} se muestran las palabras clave de autor, su cuadrante superior derecho contiene parte del tema ``mobility'' considerado como un grupo de subtemas bien desarrollados e importantes para el campo de investigación de los ``algoritmos o métodos para agrupamiento de trayectorias GPS''. 
Su cuadrante superior izquierdo cuenta con el tema ``vehicle trajectory'' y parcialmente con los temas ``mobility'' y ``urban computing'', es decir, que contienen subtemas bien desarrollados, aunque no son de importancia para el campo de investigación. Su cuadrante inferior izquierdo contiene la otra mitad del tema ``urban computing'' dentro de este tema se encuentran subtemas que están pocos desarrollados, no son tomados en cuenta, son emergentes o han desaparecidos. 
Por último, su cuadrante inferior derecho cuenta con los temas ``trajectory cluster'' y ``clustering'', es decir, contienen subtemas importantes, aunque no están del todo desarrollados.

%FIGURA 2. Diagrama estratégico de las palabras clave del autor, generado con bibliometrix. Fuente: Web of Science.

\subsection{Evolución temática de las palabras claves.}

\begin{figure}
\centering
\includegraphics[width=14cm]{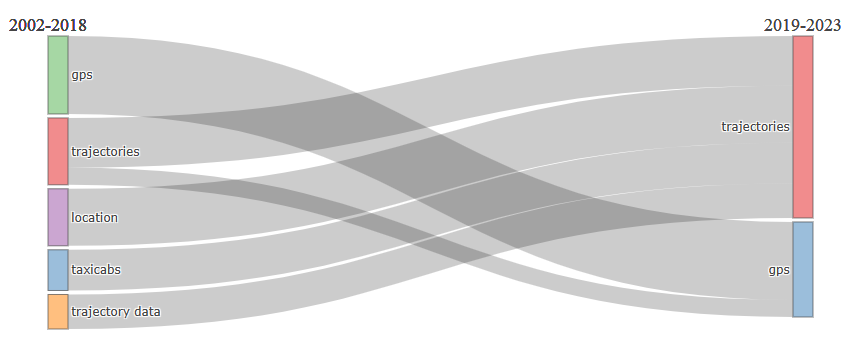}
\caption{Diagrama estratégico de las palabras clave del autor, generado con bibliometrix. Fuente: Scopus}
\label{fig:fig_3}
\end{figure}

\begin{figure}
\centering
\includegraphics[width=14cm]{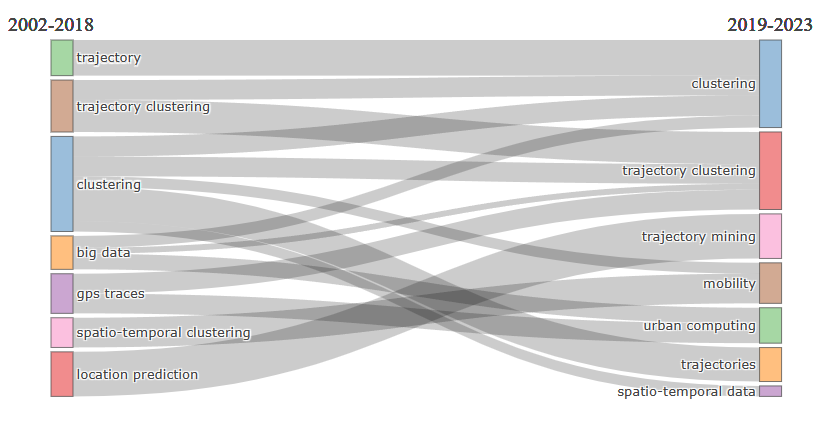}
\caption{Diagrama estratégico de las palabras clave del autor, generado con bibliometrix. Fuente: Scopus}
\label{fig:fig_4}
\end{figure}

Para el análisis temático de la evolución de las palabras claves se utilizó el paquete bibliometrix de R con su interfaz gráfica biblioshiny, en el cual, se estableció un rango de años para observar los cambios que existen entre una temática u otra. 
En la Figura~\ref{fig:fig_3} y Figura~\ref{fig:fig_4} se muestra la evolución temática de las keyWord Plus y las palabras claves del autor, desde el inicio de los estudios en el campo de investigación hasta la actualidad. 
En la Figura~\ref{fig:fig_3} el tema trajectories se mantiene, aunque se integra con algunos de los subtemas que pertenecian a data location, taxi cabs y trajectory data. Esto forma un nuevo grupo, sin embargo, prevalece el tema clustering posiblemente porque mantiene en su totalidad los subtemas que estaban presentes desde el 2002 hasta el 2018.  Los temas ``location'', ``taxi cabs'' y ``trajectory data'' también se transformaron en nuevos temas que mantienen ciertos subtemas de los temas que existían antes del 2018. Sin embargo, se observa que ``gps'' ha sufrido cambios menores en los subtemas que se han presentado hasta la actualidad.

%FIGURA 3. Diagrama estratégico de las palabras clave del autor, generado con bibliometrix. Fuente: Web of Science.

En la Figura~\ref{fig:fig_4} el tema clustering en la actualidad se ha integrado con algunos de los subtemas que pertenecían a trajectory, trajectory clustering y big data. Así mismo, el tema trajectory clustering se mantiene, aunque algunos de sus subtemas pasaron a ser parte del tema clustering. Otros temas como trajectory mining se conforma en su totalidad con los subtemas que antes del 2018 pertenecían al tema location prediction. En la actualidad han emergido temas como mobility, trajectories y spacio-temporal data cuyos subtemas se han derivado del tema clustering. Finalmente se observa que ninguno de los temas actuales han conservado los subtemas en su totalidad.

%FIGURA 4. Diagrama estratégico de las palabras clave del autor, generado con bibliometrix. Fuente: Web of Science.

\subsection{Grado de concentración de variables seleccionadas.}

En esta subsección se analizan algunas variables bibliométricas, con el fin de mostrar el grado con el que se concentran. Según \citet{stuart_open_2018} los estudios bibliométricos pueden clasificarse ampliamente como relacionales o evaluativos, ya sea ofreciendo información sobre la relación entre unidades de análisis o ayudando en la evaluación de unidades de análisis. Para realizar este tipo de análisis, se hace uso de la teoría de la información propuesta por \citet{shannon_mathematical_1948}, esta teoría proporciona diferentes métricas que permiten obtener información, como la desviación estándar, la asimetría o la curtosis. Así mismo desarrolló su propia métrica denominada la entropía de Shannon, que mediante una distribución de probabilidad discreta $P=\{p_{j};j=1,\ldots,N\}$ con $\sum_{j=1}^{N}{x_{i}p_{j}=1}$ la entropía de Shannon se define como:

\begin{equation}
S[P]=-\sum_{j=1}^{N}{p_j \ln (p_j)}
\end{equation}

La entropía de Shannon puede ser interpretada o utilizada de muchas maneras en otros campos científicos. \citet{mejia-barron_shannon_2019} hacen uso de la Entropía de Shannon y un sistema de lógica difusa para diagnosticar fallas de cortocircuitos, en otro artículo \citet{babichev_hybrid_2019} presenta la tecnología de reducción de perfiles de expresión genética basada en el uso complejo de métodos de lógica difusa, criterios estadísticos y la entropía de Shannon. Por otra parte, \citet{savakar_copy-move_2020} discuten sobre la detección de falsificación de una imagen utilizando la Entropia de Shannon y medidas de similitud y disimilitud. Finalmente, es utilizada en estudios bibliométricos con el fin de estudiar la distribución de equidad/concentración de diferentes variables importantes como temas de investigación, autores, entre otros \citep{polyakov_does_2017}.
Para una mejor interpretación de la información, se hace uso de la Entropía de Shannon en su forma normalizada, dividiéndola por su valor máximo. Por lo tanto, el índice de concentración normalizado se define de la siguiente manera:

\begin{equation}
H[P]= \frac{S[P]}{S_{MAX}} = \frac{-\sum_{j=1}^{N}{p_j \ln (p_j)}}{\ln N}
\end{equation}

Bajo la condición de $0 \leq H \geq 1$, donde $H=1$ significa que todas las categorías están representadas de manera uniforme, es decir, existe ausencia de concentración, y $H=0$ que la distribución está concentrada en un solo punto. 
Se calculó el índice de concentración entrópica normalizado para la distribución de autores, fuentes, países, áreas de investigación y citas.
Los resultados se observan en la Tabla~\ref{tab:table_9}. Donde se observa que los autores se encuentran distribuidos de manera uniforme. Las fuentes también se encuentran distribuidas de manera uniforme, como se observa en la Tabla~\ref{tab:table_5}. Los países que publican artículos relacionados con el tema de estudio están muy concentrados en unos pocos países como se muestra en la Tabla~\ref{tab:table_3}. Sin embargo, tomando en cuenta el valor del índice de autores y países se presenta que la distribución de los autores dentro de estos países se encuentra distribuidos de manera uniforme. Del mismo modo en las áreas de investigación se detectó una concentración moderadamente baja, esto puede ser observado en la Tabla~\ref{tab:table_1} donde el 74,5\% de publicaciones se reparte entre el área de ciencias de la computación, ingeniería, ciencias sociales y matemáticas. Finalmente, los artículos más citados se encuentran moderadamente concentrados como se observa en la Tabla~\ref{tab:table_6}.

% Tabla IX
\begin{table}
\caption{Índice de concentración entrópica (H) de las variables seleccionadas. Fuente: Scopus.}
\centering
\begin{tabular}{ll}
\toprule
Variable & H \\
\midrule
Autores	& 0,9665 \\
Fuentes	& 0,9211 \\
Países	& 0,5375 \\
Áreas de investigación & 0,6682 \\
Citas de artículos & 0,8169 \\
\bottomrule
\end{tabular}
\label{tab:table_9}
\end{table}

Una medida alternativa para observar la distribución que siguen los autores según su productividad es la ley de Lokta. Según el hallazgo empírico que realizó \citet{lotka_frequency_1926}, la ley de Lokta sigue una forma de ley de Zipf. El hallazgo original, fue basado en una base de datos restringidas a la física y a la química. Su ecuación en base a esta restricción se define a continuación:

\begin{equation}
a_n=\frac{a_1}{n^2},n=1,2,\ldots,N
\end{equation}

Donde $a_n$ es el número de autores que publican $n$ artículos y $a_1$ es el número de autores que publican un solo artículo.
\citet{lotka_frequency_1926} dedujo su ley empírica a partir de una muestra muy específica, sin embargo, para una generalización de su ecuación podría ser:

\begin{equation}
a_n=\frac{a_1}{n^c},n=1,2,\ldots,N
\end{equation}

donde $c$ es un parámetro que debe estimarse para que se ajusten mejor a los datos de la distribución.
El valor de $c = 2,52$, con un $R^2= 0,96$. La Tabla~\ref{tab:table_10} resume la distribución real y ajustada del número de autores que publican $n$ artículos. 
Se observa que el número real de autores que publican solo 1 artículo es menor que el predicho por la ley de Lokta confirmando que la autoría no está distribuida de manera más amplia y uniforme.

% Tabla X
\begin{table}
\caption{Distribución observada del número de autores que escribieron un número determinado de artículos y valores ajustados de la ley de Lotka. Fuente: Scopus.}
\centering
\begin{tabular}{llll}
\toprule
\# Artículos & \# Autores & Frecuencia observada & \# Frecuencia ajustada \\
\midrule
1 & 1080 & 0,7627 & 0,7630 \\
2 & 184 & 0,1299 & 0,1300 \\
3 & 64 & 0,0452 & 0,0450 \\
4 & 35 & 0,0247 & 0,0250 \\
5 & 17 & 0,0120 & 0,0120 \\
6 & 12 & 0,0085 & 0,0080 \\
7 & 6 & 0,0042 & 0,0040 \\
8 & 6 & 0,0042 & 0,0040 \\
9 & 2 & 0,0014 & 0,0010 \\
10 & 2 & 0,0014 & 0,0010 \\
11 & 4 & 0,0028 & 0,0030 \\
12 & 2 & 0,0014 & 0,0010 \\
15 & 1 & 0,0007 & 0,0010 \\
16 & 1 & 0,0007 & 0,0010 \\
\bottomrule
\end{tabular}
\label{tab:table_10}
\end{table}

\section{Gráficos de citas, fuentes y autores}

\begin{figure}
\centering
\includegraphics[width=14cm]{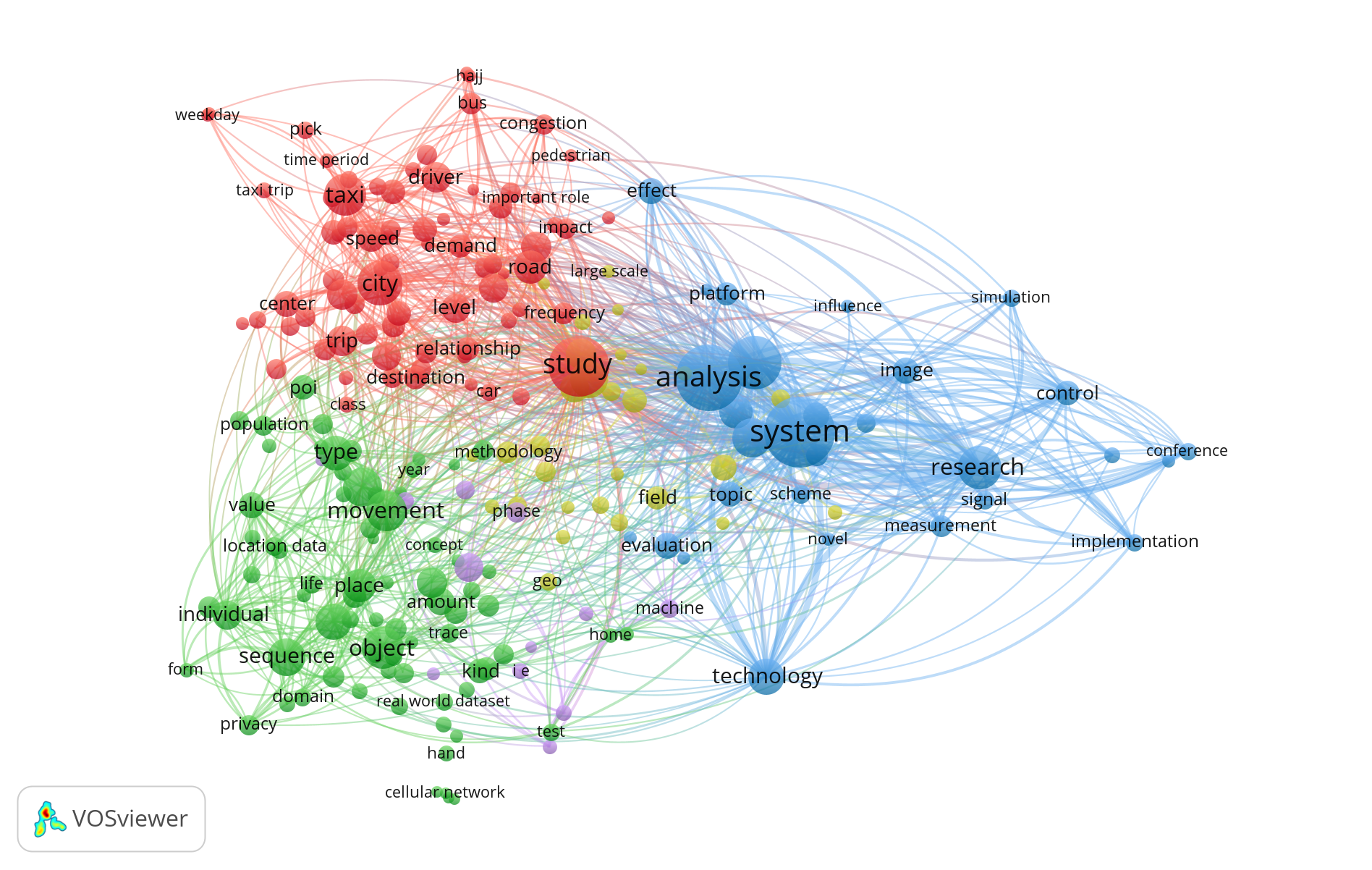}
\caption{Mapa de nubes de palabras en títulos y resúmenes (recuento completo), generado con VOSviewer. Fuente: Scopus.}
\label{fig:fig_5}
\end{figure}

Las siguientes figuras fueron generadas utilizando la herramienta de software VOSviewer que permite crear mapas basados en red, permitiendo la visualización y la exploración. Desarrollado por \citet{van_eck_software_2010} nos permite contar las palabras que aparecen en el título, el resumen y las palabras claves, obteniendo las relaciones que aparecen en los diferentes documentos que se encuentran publicados. 
La Figura~\ref{fig:fig_5} representa el mapa de nubes con las palabras que son relevantes en los artículos. El mapa muestra la cantidad de veces en que las palabras aparecen en los artículos y que tanta es la relación que existe entre ellos. 
En mapa se divide por grupos, la parte azul tiene una concentración de la palabra sistema, que a su vez se relaciona con la palabra análisis, investigación, tecnología y evaluación. En la parte roja se observan palabras que están relacionadas con la planificación urbana o urbanismo, entre sus palabras están estudio, taxi, carretera, demanda, congestión. En la parte verde, amarilla y morada hacen alusión a los conceptos asociados al movimiento de los objetos y sus diferentes aplicaciones.  Se destacan las palabras estudio, ciudad, análisis, sistema y movimiento porque crean los nexos entre todo el conjunto de palabras, esto ha permitido que se detecten nuevas perspectivas de análisis hacia aplicaciones emergentes como la propuesta por \citet{reyes_proposal_2022}.

%FIGURA 5. Mapa de nubes de palabras en títulos y resúmenes (recuento completo), generado con VOSviewer. Fuente: Web of Science.

\begin{figure}
\centering
\includegraphics[width=14cm]{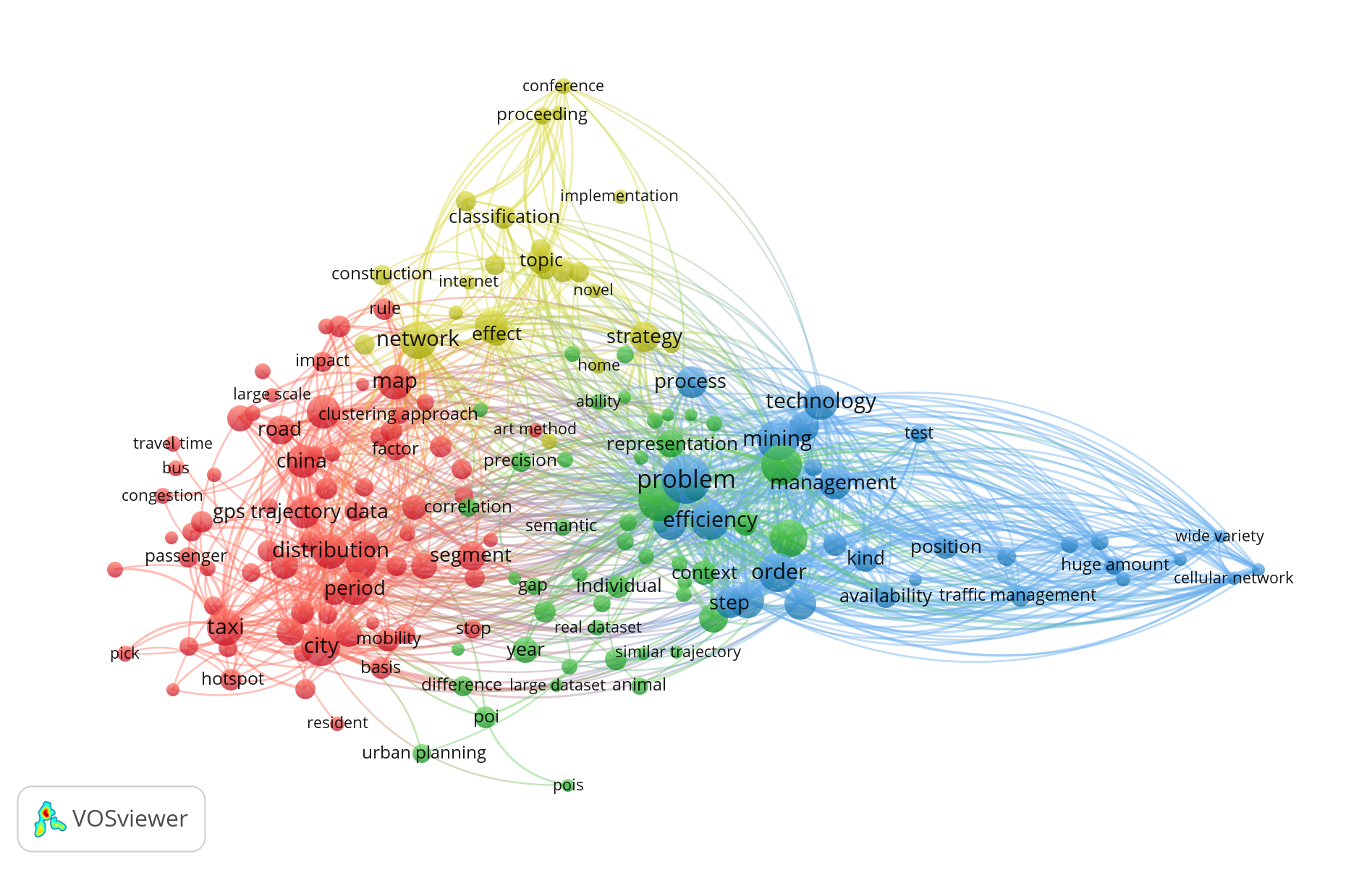}
\caption{Mapa de nubes de palabras en títulos y resúmenes (recuento binario), generado con VOSviewer. Fuente: Scopus}
\label{fig:fig_6}
\end{figure}

La Figura~\ref{fig:fig_6} es una representación casi similar a la Figura~\ref{fig:fig_5}, con la diferencia de que las palabras se cuentan de forma binaria. Esto significa que cuando aparece una palabra, Vos-Viewer solo la va a contar una vez independientemente del número de veces que aparezca en el documento. Esta ligera diferencia puede cambiar los resultados que se obtuvieron con las gráficas anteriores, porque si una palabra se repite mucho no entra en el conteo del resultado final.
En el mapa de nubes se observa que la parte amarilla de la Figura~\ref{fig:fig_6} se han fusionado con las palabras que tenían que ver con clasificación, tema, estrategia y redes, siendo la principal diferencia entre la Figura~\ref{fig:fig_5} y la Figura~\ref{fig:fig_6}. Sin embargo, aún está presente la parte roja con las palabras que hacen alusión a conceptos asociados a la planificación urbana y sus diferentes aplicaciones, también se mantiene la parte azul y verde con temas muy relacionados a la gestión y eficiencia de problemas derivados del urbanismo.

%FIGURA 6. Mapa de nubes de palabras en títulos y resúmenes (recuento binario), generado con VOSviewer. Fuente: Web of Science.

\begin{figure}
\centering
\includegraphics[width=14cm]{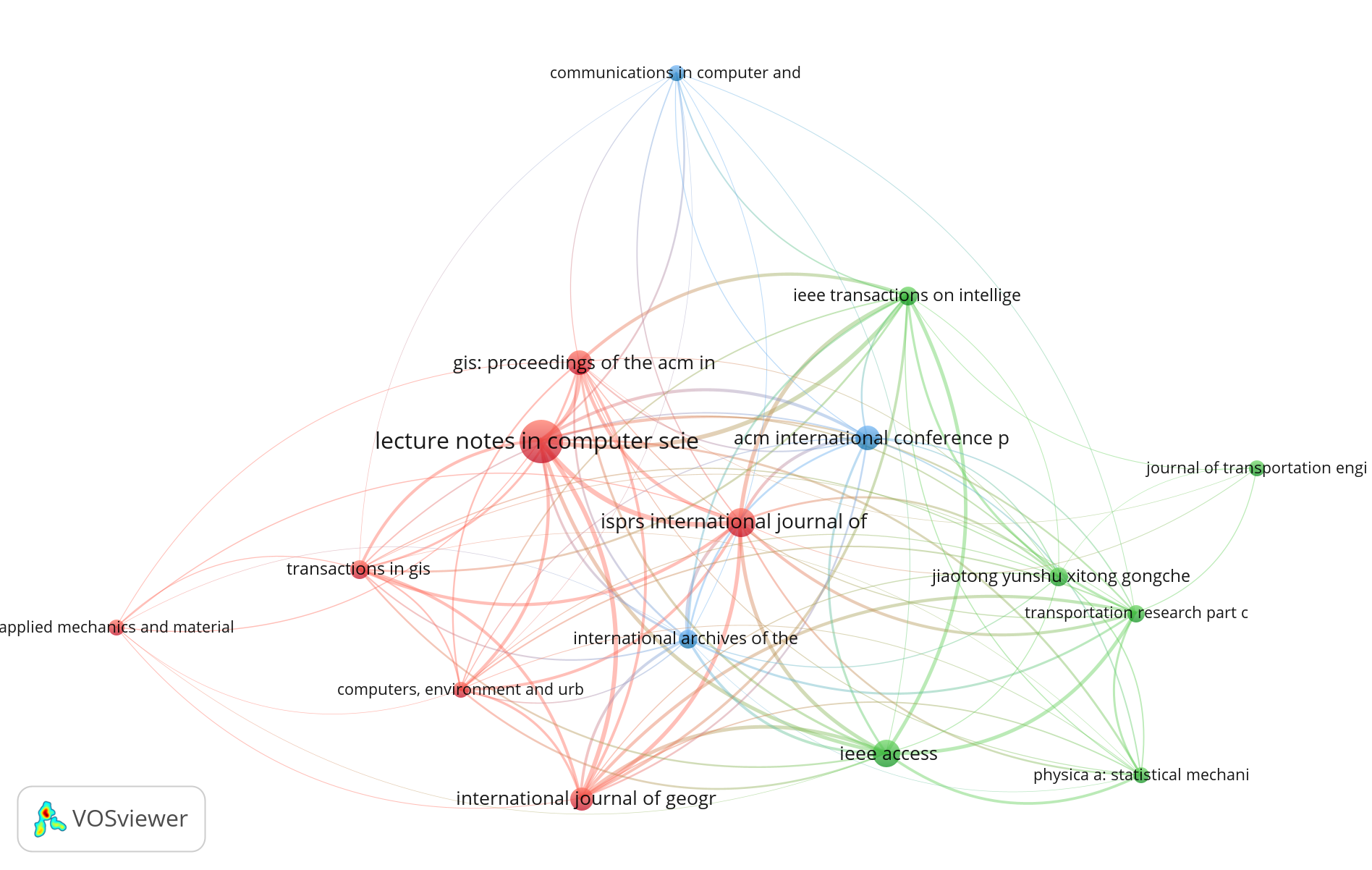}
\caption{Mapa en la nube de revistas donde se publicaron artículos sobre ``agrupamiento de trayectorias GPS'', generado con VOSviewer. Fuente: Scopus}
\label{fig:fig_7}
\end{figure}

La Figura~\ref{fig:fig_7} muestra el mapa de nubes de las fuentes de los artículos. En el mapa se diferencian las revistas, que hacen referencia a la Tabla~\ref{tab:table_5} cada una de las fuentes publica artículos relacionados con algoritmos o métodos de agrupamiento de trayectorias, agrupamiento de trayectorias GPS, urbanismo, planificación, tráfico, entre otras.

%FIGURA 7. Mapa en la nube de revistas donde se publicaron artículos sobre ``agrupamiento de trayectorias GPS'', generado con VOSviewer. Fuente: Web of Science.

\begin{figure}
\centering
\includegraphics[width=14cm]{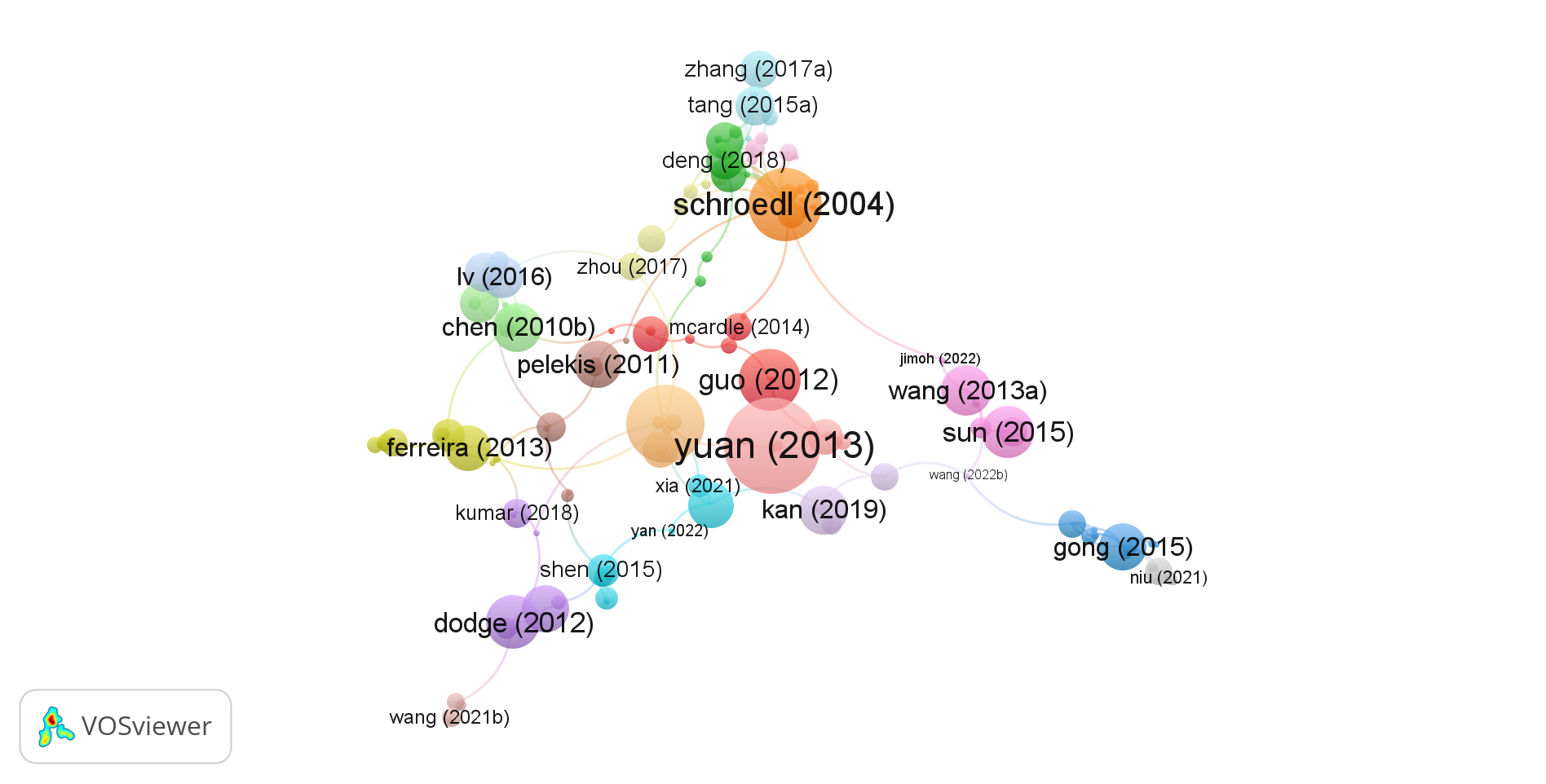}
\caption{Se publicaron mapas de nubes de revistas de autores con artículos sobre ``agrupamiento de trayectorias GPS'', generado con VOSviewer. Fuente: Scopus}
\label{fig:fig_8}
\end{figure}

La Figura~\ref{fig:fig_8} muestra todos los artículos que pertenecen a la muestra y el tamaño del nodo que se crea depende del número de citas que tienen. Este resultado se puede observar en la Tabla~\ref{tab:table_6} de la subsección de artículos más citados.
En la Figura~\ref{fig:fig_8} se observa que los dos nodos que más destacan son el de \citet{jing_yuan_t-drive_2013} publicado en IEEE Xplore con el diseño de un enfoque de agrupación basada en varianza-entropía para la estimación en la distribución de los tiempos de viajes entre dos puntos diferentes, y \citet{schroedl_mining_2004} publicado en  Data Mining and Knowledge Discovery que presentan un enfoque para inducir mapas de alta precisión a partir de trazas de vehículos equipados con receptores GPS diferenciales. 

También se destacan otros autores por su cantidad de citas como \citet{tang_uncovering_2015} publicado en Physica A: Statistical Mechanics and its Applications o \citet{guo_discovering_2012} publicado en Transactions in GIS.

%FIGURA 8. Se publicaron mapas de nubes de revistas de autores con artículos sobre ``agrupamiento de trayectorias GPS'', generado con VOSviewer. Fuente: Web of Science.

\section{Posibles líneas de investigación}

Aunque existen una gran variedad de algoritmos de agrupamiento para trayectorias, hay muy pocas revisiones bibliográficas o de literatura acerca del funcionamiento o a que campos de investigación pueden ir dirigidos, el único que trata limitadamente es \citet{yuan_review_2017} con su análisis de los algoritmos de agrupamiento para trayectorias, sin embargo, él lo adapta a un contexto general del tema de estudio. 
Según la revisión bibliométrica el estudio de los datos GPS que se obtienen a través de los vehículos puede ayudar a resolver problemas tanto viales como urbanos, por lo tanto, en esta línea aún faltan estudios que entreguen pautas de inicio a nuevos investigadores que deseen incurrir en el campo de agrupamiento de trayectorias GPS como por ejemplo, identificar que vías, espacios aéreos o marítimos son los más adecuados para la rápida movilidad de los medios de transporte multimodales, planificar las rutas de urbanizaciones modernas o que se encuentren en fase de construcción para disminuir el tráfico vehicular, analizar que patrones son los que provocan los accidentes de tránsito para así tratar de evitarlos, determinar que rutas son las más factibles para que circulen los vehículos autónomos, establecer caminos, calles o carreteras seguras para las personas que usen un medio de transporte distinto como bicicletas, scooters, patines, entre otros. 
Finalmente, también se puede incurrir en la revisión de los algoritmos de agrupamiento de trayectorias centrados en otro ámbito, tanto en el análisis de la movilidad o migración de animales, personas, trayectorias de robots, vehículos aéreos no tripulados, análisis de trayectorias de huracanes entre otros. En relación con este aspecto, casi no hay documentos que indiquen la tendencia que llevan los algoritmos o métodos de agrupamiento para trayectorias GPS en este campo de investigación.

\section{Conclusión}

Este análisis muestra que el agrupamiento para trayectorias GPS comprenden una combinación entre el urbanismo y los efectos que tienen los vehículos en las calles, caminos o carretas. Cabe destacar que esto no sería posible sin Sistemas de posicionamiento global o GPS. Además de la integración de los algoritmos de agrupamiento de trayectorias correctos ya sea TraClus, Kmeans, Tra-Dbscan entre otros. 
Este artículo constituye un aporte significativo al análisis bibliométrico acerca de los algoritmos o métodos de agrupamiento para trayectorias GPS, que comprende 559 artículos publicados en Web of Science, estos registros permitieron encontrar resultados significativos, tanto en las relaciones que tienen entre palabras claves, autores, citas, entre otros. 
Se detecto que existen artículos importantes que no se encuentran en Scopus, por ejemplo, Time-focused clustering of trajectories of moving objects de \citet{nanni_time-focused_2006}, considerado en otras fuentes bibliográficas como un artículo muy citado. 
En la Tabla~\ref{tab:table_9} se observa una alta concentración de autores de China, aunque la diversidad de países si predomina. Además, se observa en la Figura~\ref{fig:fig_8} que las citas realizadas entre artículos se encuentran muy relacionadas, indicando posiblemente que el tema de estudio se está consolidando.

% \section*{Acknowledgments}
% This was was supported in part by......

%Bibliography
% \bibliographystyle{unsrt}
\bibliographystyle{plainnat}
\bibliography{article}

\end{document}